\documentclass{mem}
\usepackage{natbib}\usepackage{txfonts}\usepackage{balance}
\usepackage{graphicx}
\usepackage[a4paper]{hyperref}
\idline{00}{001}
\begin{document}
\def\teff{$T\rm_{eff }$}
\def\kms{$\mathrm {km s}^{-1}$}

\title{
VLTI observations of AGB stars
}

\author{
M. \,Wittkowski\inst{1} \and
D.~A.~Boboltz\inst{2}\and
T.~Driebe\inst{3}\and
K.~Ohnaka\inst{3}}
\offprints{M. Wittkowski, Email: mwittkow@eso.org}

\institute{
European Southern Observatory, Garching, Germany
\and U.S. Naval Observatory, Washington, DC, USA 
\and Max-Planck-Institut f\"ur Radioastronomie, Bonn, 
Germany}
\authorrunning{Wittkowski et al.}
\titlerunning{VLTI observations}
\abstract{We report on recent observations of AGB stars
obtained with the VLT Interferometer (VLTI).
We illustrate in general the potential of interferometric 
measurements to study stellar atmospheres and circumstellar envelopes, 
and demonstrate 
in particular the advantages of a coordinated multi-wavelength approach 
including near/mid-infrared as well as radio interferometry. 
We report on studies of the atmospheric structure of non-Mira and
Mira variable giants.
We have used VLTI observations of the near- and mid-infrared stellar sizes 
and concurrent VLBA observations of the SiO maser 
emission. So far, this project includes studies of the Mira
stars S\,Ori and RR\,Aql as well as of the supergiant AH\,Sco. 
The results from our first epochs of S\,Ori measurements have recently 
been published and the main results 
are reviewed here. The S\,Ori maser ring is found to lie at a mean distance
of approximately 2 stellar radii, a result that is virtually free of the 
usual uncertainty inherent in comparing observations of variable stars 
widely separated in time and stellar phase. We discuss the status of our 
more recent S\,Ori, RR\,Aql, and AH\,Sco observations, and present an 
outlook on the continuation of our project.
}
\maketitle{}
\section{Introduction}
The evolution of cool luminous stars,
including Mira variables, is accompanied by significant mass-loss to the
circumstellar environment (CSE) with mass-loss rates of up
to $10^{-7} - 10^{-4}$\,M$_\odot$/year.
The detailed structure of the CSE, the detailed physical nature of this
mass-loss process from evolved stars, and especially
its connection with the pulsation mechanism in the case of Mira
variable stars, are not well understood.
Furthermore, one of the basic unknowns in the study of late-type stars
is the mechanism by which usually spherically symmetric stars on the 
asymptotic giant branch (AGB) evolve to form axisymmetric or bipolar
planetary nebulae (PNe).  There is evidence for some 
asymmetric structures already at the AGB or supergiant stage
\citep[e.g.,][]{mwi:weigelt96,mwi:weigelt98,mwi:wittkowski98,mwi:monnier99}.

\begin{figure}[tb]
\centering
\includegraphics[width=0.47\textwidth]{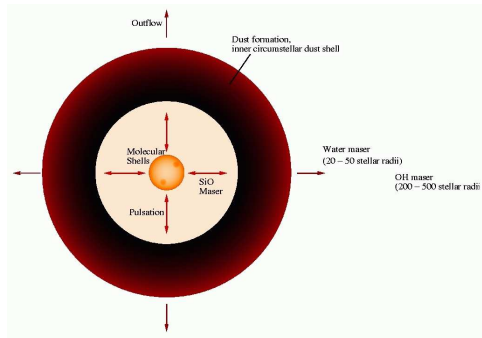}
\caption{Sketch of a Mira variable star and its circumstellar
envelope (CSE). A multi-wavelength study (MIDI/AMBER/VLBA) is well suited
to probe the different regions shown here. Owing to the stellar variability,
only contemporaneous observations are meaningful.}
\label{mwi:scheme}
\end{figure}
Coordinated multi-wavelength studies (near-infrared, mid-infrared,
radio, millimeter) of the stellar surface (photosphere) {\it and} the CSE 
at different
distances from the stellar photosphere and obtained at corresponding 
cycle/phase values of the stellar variability curve are best suited to
improve our general understanding of the atmospheric structure, the CSE, 
the mass-loss process, and ultimately of the evolution of 
symmetric AGB stars toward axisymmetric or bipolar planetary nebulae. 
Fig.~\ref{mwi:scheme} shows a schematic view of a Mira variable star, 
indicating the different regions that can be probed by different 
techniques/wavelength ranges (VLTI/AMBER, VLTI/MIDI, VLBA/maser, ALMA).

The conditions near the stellar surface can best be studied
by means of optical/near-infrared long-baseline interferometry
\citep[see, e.g.,][]{mwi:haniff95,mwi:wittkowski01,mwi:thompson02,
mwi:wittkowski04,mwi:woodruff04,mwi:boboltz05,mwi:fedele05}.
The structure and physical parameters of the molecular shells located
between the photosphere and the dust formation zone, as well as of the
dust shell itself can be probed by mid-infrared interferometry 
\citep[e.g.,][]{mwi:danchi94,mwi:ohnaka05}.

Complementary information regarding the molecular shells can be
obtained by observing the maser radiation that some of these molecules
emit. The structure and dynamics of the CSE of Mira variables and other
evolved stars has been investigated by mapping SiO maser emission
at typically about 2 stellar radii toward these stars using
very long baseline interferometry (VLBI) at radio wavelengths
\citep[e.g.,][]{mwi:boboltz97,mwi:kemball97,mwi:boboltz05}.

Results regarding the relationships between the different regions
mentioned above and shown in Fig.~\ref{mwi:scheme} suffer often from 
uncertainties inherent in comparing observations of variable stars widely 
separated in time and stellar phase 
\citep[see the discussion in][]{mwi:boboltz05}. Both, the
photospheric stellar size, as well as the mean diameter of the
SiO maser shell are known to vary as a function of the 
stellar variability phase with amplitudes of 20-50\% 
\citep{mwi:ireland04,mwi:thompson02,mwi:humphreys02,mwi:diamond03}.
  
To overcome these limitations, we have established
a program of coordinated and concurrent observations at near-infrared, 
mid-infrared, and radio wavelengths of evolved stars, aiming at a better 
understanding of the structure of the CSE, of the mass-loss process, and of
the triggering and formation of asymmetric structures. 
\section{The atmospheric structure of non-Mira and Mira giants}
Fundamental parameters, most importantly radii and effective temperatures,
of regular cool giant stars have frequently been obtained
with interferometric and other high angular resolution techniques,
thanks to the favorable brightness and size of these stars.
Further parameters of the stellar structure, as the strength
of the limb-darkening effect, can be studied when more than one resolution
element across the stellar disk is employed.
Through the direct measurement of the center-to-limb intensity 
variation (CLV) across stellar disks and their close environments, 
interferometry probes the vertical temperature profile, as well as 
horizontal inhomogeneities. 
Such direct limb-darkening studies have been accomplished for
a relatively small number of stars using different interferometric
facilities \citep[including, for instance,][]{mwi:hanbury73,mwi:quirrenbach96,
mwi:hajian98,mwi:wittkowski01,mwi:wittkowski04}.

\begin{figure}[tb]
\centering
\includegraphics[width=0.23\textwidth]{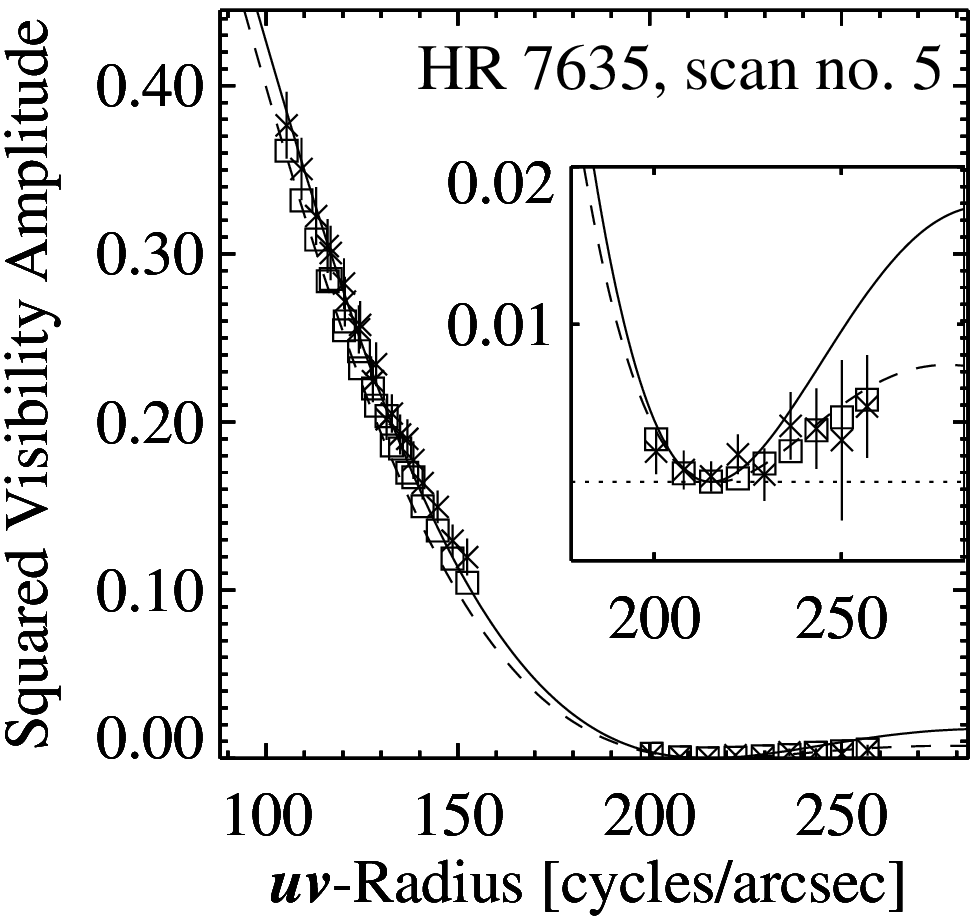}
\includegraphics[width=0.23\textwidth]{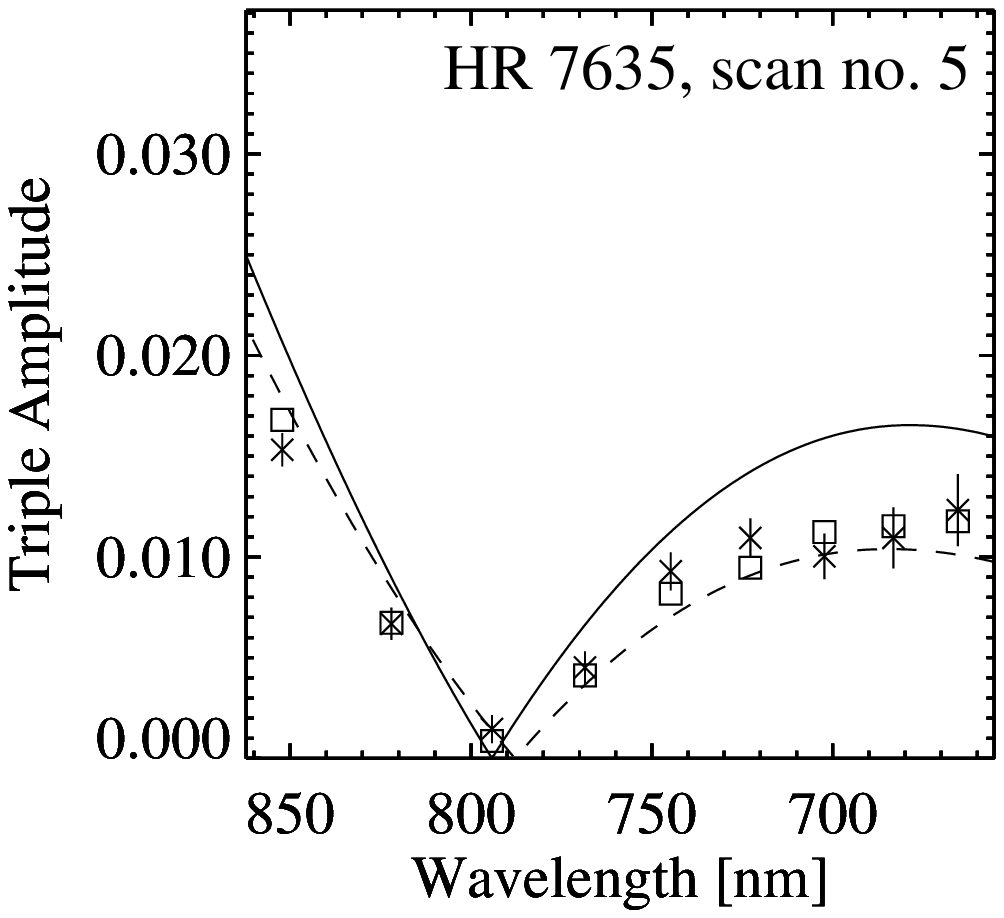}
\includegraphics[width=0.23\textwidth]{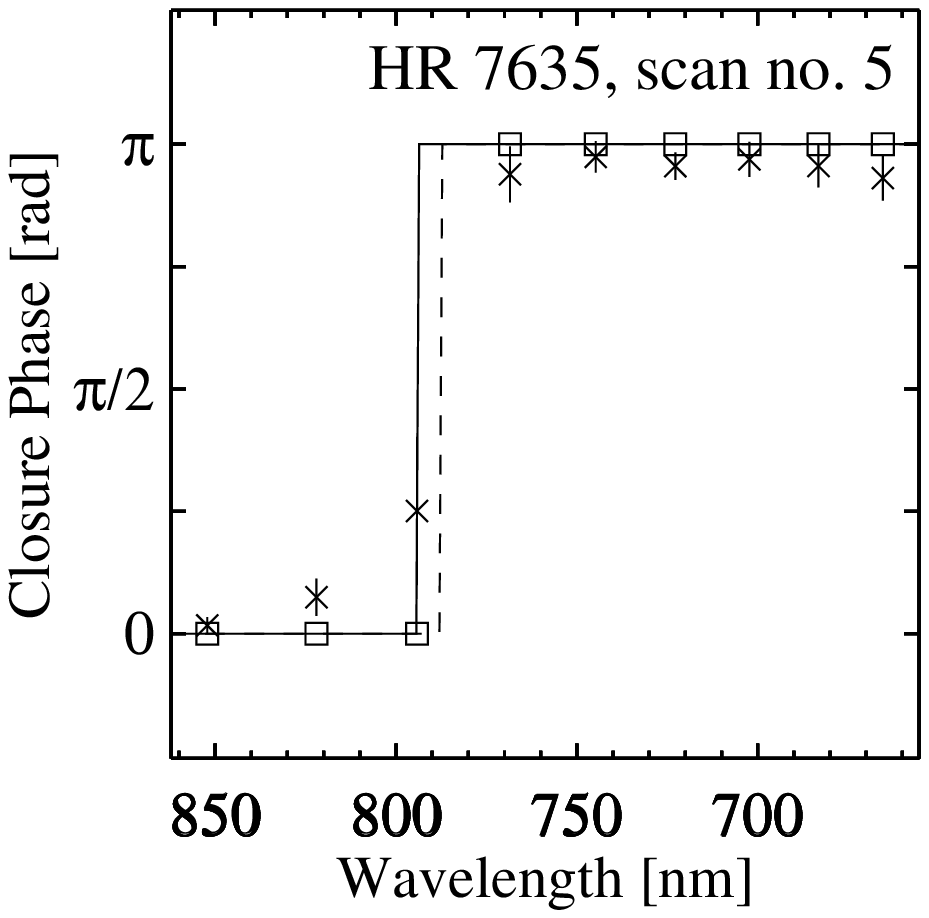}
\caption{NPOI limb-darkening observations (squared visibility amplitude,
triple amplitude, closure phase) of the M0 giant $\gamma$\,Sge, together
with a comparison to the best fitting {\tt ATLAS\,9} model atmosphere 
prediction (squares).
For comparison, the solid line denotes a uniform disk model, and the
dashed line a fully-darkened disk model. {\tt ATLAS\,9} models with
variations of $T_\mathrm{eff}$ and $\log g$ result in significantly different
model predictions. From \citet{mwi:wittkowski01}.}
\label{mwi:gamsge}
\end{figure}
Recent optical multi-wavelength measurements
of the cool giants $\gamma$\,Sge and BY\,Boo
\citep{mwi:wittkowski01} succeeded not only in
directly detecting the limb-darkening
effect, but also in constraining {\tt ATLAS\,9} \citep{mwi:kurucz93}
model atmosphere parameters. Fig.~\ref{mwi:gamsge} shows
one dataset including squared visibility amplitudes, triple amplitudes,
and closure phases of the M0 giant $\gamma$\,Sge obtained with NPOI,
together with a comparison to the best fitting {\tt ATLAS\,9} model 
atmosphere prediction. {\tt ATLAS\,9} models with
variations of $T_\mathrm{eff}$ and $\log g$ result in significantly different
model predictions. By this direct comparison of the NPOI data to the
{\tt ATLAS 9} models alone, the effective temperature of $\gamma$\,Sge
is constrained to 4160$\pm$100\,K. The limb-darkening observations are 
less sensitive to variations of the surface gravity, and $\log g$ 
is constrained to 0.9$\pm$1.0 \citep{mwi:wittkowski01}. These results 
are well consistent with independent estimates, such as calibrations
of the spectral type. 

\begin{figure}[tb]
\centering
\includegraphics[width=0.4\textwidth]{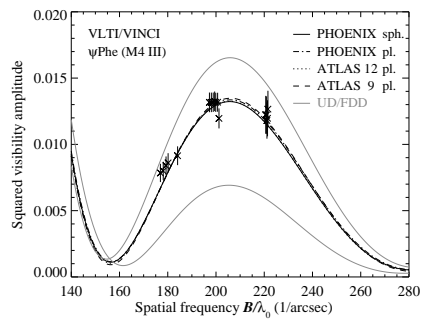}
\caption{VLTI limb-darkening observations of the M4 giant $\psi$\,Phe
\citep{mwi:wittkowski04}.}
\label{mwi:psiphe}
\end{figure}
The first
limb-darkening observation obtained with the VLTI succeeded
in the early commissioning phase of the VLTI \citep{mwi:wittkowski04}.
Using the VINCI instrument, $K$-band visibilities of the
M4 giant $\psi$ Phe were measured in the first and second lobe of the 
visibility function.
These observations were found to be consistent with predictions
by {\tt PHOENIX} and {\tt ATLAS} model atmospheres, the parameters for
which were constrained by comparison to available spectrophotometry and
theoretical stellar evolutionary tracks (see Fig.~\ref{mwi:psiphe}). 
Such limb-darkening
observations also result in very precise and accurate
radius estimates because of the precise description of the CLV. Future use
of the spectro-interferometric capabilities of AMBER and MIDI will
enable us to study the wavelength-dependence of the limb-darkening effect,
which results in stronger tests and constraints of the model atmospheres
than these broad-band observations (cf. the wavelength-dependent optical
studies with NPOI as described above).

For cool pulsating Mira stars, the CLVs are expected to be more complex
than for non-pulsating M giants due to the effects of molecular layers
close to the continuum-forming layers. Self-excited hydrodynamic model 
atmospheres of Mira stars have been presented \citep[][Scholz \& Wood, 
private communication]{mwi:hofmann98,
mwi:tej03,mwi:ireland04}.  Different radius definitions, such
as the Rosseland mean radius, the continuum radius, or the radius at which 
the filter-averaged intensity drops by 50\%, may result in different
values for the same CLV. 
On these topics, see also \citet{mwi:scholz03}.
However, interferometric measurements covering a range of spatial frequencies
can directly be compared to CLV predictions by these model atmospheres
without the need of a particular radius definition. At pre-maximum
stellar phases, when the temperature is highest, the broad-band CLVs are 
less contaminated by molecular layers, and different radius definitions
agree relatively well (Scholz \& Wood, private communication).

$K$-band VINCI observations of the prototype
Mira stars  $o$\,Cet and R\,Leo have been presented by \citet{mwi:woodruff04}
and \citet{mwi:fedele05}, respectively.
These measurements at post-maximum stellar phases
indicate indeed $K$-band CLVs which are clearly different from a
uniform disk profile already in the first lobe of the visibility
function. The measured visibility values were found to
be consistent with predictions by the self-excited dynamic
Mira model atmospheres described above that include molecular
shells close to continuum-forming layers.
\section{Joint VLTI/VLBA observations of the Mira star S\,Ori}
\begin{figure}[tb]
\centering
\includegraphics[width=0.5\textwidth]{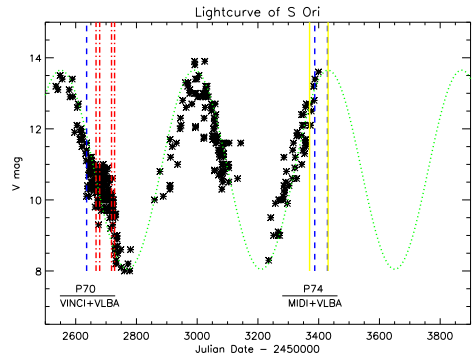}
\caption{Lightcurve of S\,Ori together with the epochs of
our joint VLTI/VLBA measurements obtained so far. Note that the y-axis 
is given with increasing $V$ magnitude, i.e. the stellar maximum is at the 
bottom and stellar minimum at the top. The study of S\,Ori was started
in ESO period P70 (Dec. 2002/Jan. 2003) including near-infrared
$K$-band VINCI and VLBA/SiO maser observations \citep{mwi:boboltz05}.
In December 2004/January 2005, we obtained concurrent mid-infrared 
VLTI/MIDI and VLBA/SiO maser observations.}
\label{mwi:lightcurve}
\end{figure}
\begin{figure}[tb]
\centering
\includegraphics[width=0.46\textwidth]{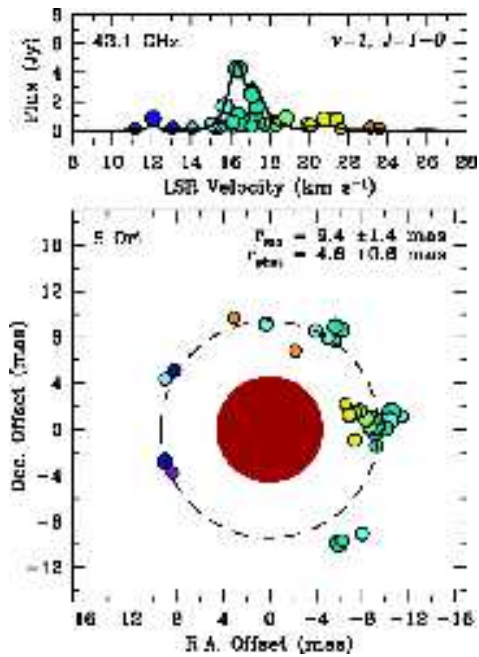}
\caption{First-ever coordinated observations between ESO's VLTI and
NRAO's VLBA facilities: 43.1\,GHz SiO maser emission toward the Mira 
variable S~Ori measured with the VLBA,
together with the near-infrared diameter measured quasi simultaneously
with the VLTI (red stellar disk).
From \citet{mwi:boboltz05}.}
\label{mwi:sori}
\end{figure}
We started our project of joint VLTI/VLBA observations of Mira stars
in December 2002/January 2003 with coordinated near-infrared 
$K$-band VLTI/VINCI observations of the stellar diameter of the Mira 
variable S\,Ori and quasi-simultaneous VLBA observations of the 43.1\,GHz 
and 42.8\,GHz SiO maser emissions toward this star \citep{mwi:boboltz05}.
We obtained in December 2004/January 2005 further concurrent observations
including mid-infrared VLTI/MIDI observations to probe the molecular
layers and the dust shell of S\,Ori, and new epochs of VLBA observations
of the 43.1\,GHz and 42.8 GHz SiO maser rings.
 
The December 2002/January 2003 represent the first-ever coordinated 
observations between the VLTI and VLBA facilities, and the results from 
these observations were recently published in \citet{mwi:boboltz05}.
Analysis of the SiO maser data recorded at
a visual variability phase 0.73 show the average distance of the masers
from the center of the distribution to be 9.4~mas for the 
$v=1, J=1-0$ (43.1 GHz) masers and 8.8~mas for the $v=2, J=1-0$ (42.8 GHz) 
masers. The velocity structure of the
SiO masers appears to be random with no significant indication of
global expansion/infall or rotation.
The determined near-infrared, $K$-band, uniform disk (UD) diameters
decreased from $\sim$\,10.5\,mas at phase 0.80 to $\sim$10.2\,mas at
phase 0.95.  For the epoch of our VLBA measurements,
an extrapolated UD diameter of $\Theta_\mathrm{UD}^K=10.8 \pm 0.3$\,mas
was obtained, corresponding to a linear radius
of $R_\mathrm{UD}^K = 2.3 \pm 0.5$~AU or
$R_\mathrm{UD}^K =490 \pm 115~R_\odot$. The model predicted difference 
between the continuum and $K$-band UD diameters is relatively low in the 
pre-maximum region of the visual variability curve as in the case of our 
observations (see above). At this phase of 0.73, the continuum diameter may 
be smaller than the $K$-band UD diameter by about 15\% \citep{mwi:ireland04}.
With this assumption, the continuum photospheric diameter
for the epoch of our VLBA observation would be
$\Theta_\mathrm{Phot}(\mathrm{VLBA\ epoch,\ phase=0.73})\approx 9.2$\,mas.
Our coordinated VLBA/VLTI measurements show that the masers lie
relatively close to the stellar photosphere at a distance of $\sim$\,2
photospheric radii, consistent with model estimates by \citet{mwi:humphreys02}
and observations of other Mira stars by \citet{mwi:cotton04}.\\
The new 2004/2005 VLTI and VLBA data are currently being reduced and
analyzed.
\section{Outlook}
We concentrate on a few stars in order to understand the CSE for a
few sources in depth. In addition to the S~Ori data described
above, we have to date VLTI/MIDI observations of the supergiant
AH~Sco, and of the Mira star 
RR~Aql, as well as concurrent VLBA observations for 
each of these targets/epochs. These data are currently being analyzed.

Further studies will aim at including more detailed near-infrared studies
of the stellar atmospheric structure (close to the photosphere) employing
VLTI/AMBER, concurrent with VLTI/MIDI and VLBA observations as
discussed above. Making use of the spectro-interferometric capabilities
of AMBER, and of the closure-phase information, these studies can
in principle reveal horizontal surface inhomogeneities 
\citep[see][]{mwi:wittkowski02}.
\bibliographystyle{aa}

\end{document}